\documentclass[nologo,11pt,a4paper]{ETHpaper}

\usepackage{amsmath}
\usepackage{amssymb}
\usepackage{amsfonts}
\usepackage{graphics}
\usepackage{graphicx}
\usepackage{subfigure}
\usepackage{array}
\usepackage{booktabs}
\usepackage{hyphenat}
\usepackage{url}

\begin{document}
\title{Higher-Order Aggregate Networks in the Analysis of Temporal Networks: Path structures and centralities}
\titlealternative{Higher-Order Aggregate Networks in the Analysis of Temporal Networks}
\author{Ingo Scholtes, Nicolas Wider, Antonios Garas}
\address{Chair of Systems Design, ETH Zurich, Switzerland\\
  \url{www.sg.ethz.ch}}
\www{\url{http://www.sg.ethz.ch}}
\makeframing
\maketitle

\abstract{
Recent research on temporal networks has highlighted the limitations of a static network perspective for our understanding of complex systems with dynamic topologies.
In particular, recent works have shown that i) the specific order in which links occur in real-world temporal networks affects causality structures and thus the evolution of dynamical processes, and ii) higher-order aggregate representations of temporal networks can be used to analytically study the effect of these order correlations on dynamical processes.
In this article we analyze the effect of order correlations on path-based centrality measures in real-world temporal networks.
Analyzing temporal equivalents of betweenness, closeness and reach centrality in six empirical temporal networks, we first show that an analysis of the commonly used static, time-aggregated representation can give misleading results about the actual importance of nodes.
We further study higher-order time-aggregated networks, a recently proposed generalization of the commonly applied static, time-aggregated representation of temporal networks.
Here, we particularly define path-based centrality measures based on second-order aggregate networks, empirically validating that node centralities calculated in this way better capture the true temporal centralities of nodes than node centralities calculated based on the commonly used static (first-order) representation.
Apart from providing a simple and practical method for the approximation of path-based centralities in temporal networks, our results highlight interesting perspectives for the use of higher-order aggregate networks in the analysis of time-stamped network data.

\section{Introduction}
\label{intro}


The network perspective has provided valuable insights into the structure and dynamics of numerous complex systems in nature, society and technology.
However, most of the complex systems studied from this perspective are not static, but rather exhibit time-varying interaction topologies in which elements are only linked to each other at specific times or during particular time intervals.
While the \emph{topological} characteristics resulting from which elements are linked to which other elements have been studied extensively, the importance of the additional \emph{temporal} dimension resulting from \emph{when} these links occur has become clear only recently.
And despite an increasing volume of research, its full impact on the properties of complex systems and on the evolution of dynamical processes still eludes our understanding~\cite{Holme2012}.

\newpage

Addressing this open issue, different strands of research have focused on the question how different types of temporal characteristics of complex networked systems -- such as the activation times of nodes, the inter-event times between links, the duration and/or concurrency of interactions, or the order in which these interactions occur -- affect the properties of temporal networks as well as dynamical processes evolving on them.
For a couple of systems, it was shown that inter-event times follow heavy-tailed distributions which in turn significantly influence the speed of processes like spreading and diffusion~\cite{Iribarren2009,Karsai2011,Rocha2011,Starnini2012,Perra2012,Perra2012a,Hoffmann2013,Takaguchi2013,Rocha2013b,Karsai2014,Jo2014}.

Apart from the \emph{timing} of interactions, the \emph{order} in which these interactions occur is another important characteristic of temporal networks.
Not only does the ordering of interactions crucially affect causality in temporal networks, it has also been shown to dramatically shift the evolution of dynamical processes compared to what we would expect based on a static, time-aggregated perspective~\cite{Lentz2013,Pfitzner2013,Scholtes2014,Rosvall2014,Lambiotte2014}.
Some of these works have further taken a modeling perspective, highlighting that real-world temporal network data exhibit non-Markovian characteristics in the sequence of links which are not in line with the Markovianity assumption that is (implicitly) made when studying static representations of time-varying complex networks.
Neglecting these non-Markovian characteristics not only leads to wrong results about dynamical processes, it also leads to wrong centrality-based rankings of nodes, as well as misleading results about community structures~\cite{Scholtes2014,Rosvall2014}.

The main reason why an analysis of static, time-aggregated networks yields misleading results about the properties of temporal networks is that the ordering of links can alter path structures in temporal networks compared to what we would expect based on their static topology.
Precisely, in static network the presence of two links $(a,b)$ and $(b,c)$ connecting nodes $a$ to $b$ and $b$ to $c$ respectively necessarily imply that a path from $a$ via $b$ to $c$ exists.
However in a temporal network, for $a$ to be able to influence $c$ the link $(a,b)$ must occur \emph{before} the link $(b,c)$ and thus the presence of a path depends on the ordering of links.
This simple example highlights that the mere ordering of links in temporal networks can introduce an additional temporal-topological dimension that can neither be understood from the analysis of static, time-aggregated representations, nor from the analysis of inter-event times or node activity distributions~\cite{Pfitzner2013}.

Highlighting the important consequences introduced by the specific ordering of links in real-world temporal networks, in this article we study how this ordering affects path-based centrality measures in temporal networks.
The main contributions of our work can be summarized as follows:
\begin{enumerate}
    \item Building on the concept of time-respecting paths with a maximum time difference between consecutive links as previously discussed in~\cite{Pan2011,Holme2012}, we introduce three different notions of path-based temporal node centralities which emphasize the additional \emph{temporal-topological} dimension that is introduced due to the ordering of links in temporal networks.
        In particular, we formally define temporal variations of \emph{betweenness}, \emph{closeness} and \emph{reach} centrality and demonstrate how they can be computed based on the topology of shortest time-respecting paths emerging in temporal networks.
    \item Calculating these temporal centrality measures for six empirical data sets, we quantify to what extent a ranking of nodes based on temporal centralities coincides with a ranking of nodes based on the same measures, however calculated based on the corresponding static, time-aggregated networks.
        From our results we conclude that, possibly due to non-Markovian characteristics previously highlighted in~\cite{Pfitzner2013,Scholtes2014}, a static analysis of node centralities yields misleading results about the importance of nodes with respect to time-respecting paths.
    \item Generalizing the usual time-aggregated static perspective on temporal networks, we further develop the second-order time-aggregated representations introduced in~\cite{Scholtes2014}, obtaining higher-order time-aggregated representations which can be conveniently analyzed using standard network-analytic methods.
        Notably, despite being static representations of temporal networks, we show that these higher-order representations allow to incorporate those order correlations that have been shown to influence the causal topologies of temporal networks.
    \item We finally define generalizations of static betweenness, closeness and reach centrality based on a second-order aggregate representation of temporal networks.
        Using six data sets on temporal networks, we show that these second-order generalizations of centralities constitute highly accurate approximations for the true temporal centrality of nodes calculated based on the detailed time-respecting path structures in temporal networks.
\end{enumerate}
The remainder of this article is structured as follows:
In section~\ref{sec:definitions} we first introduce basic concepts such as our notion of temporal networks, time-aggregated and time-unfolded representations of temporal networks, as well as time-respecting paths with maximum time differences between consecutive links.
In section~\ref{sec:higherOrder} we introduce the framework of higher-order time-aggregated networks, a simple abstraction of temporal networks that takes into account the statistics of time-respecting paths up to a given length.
In section~\ref{sec:centrality} we finally define three temporal centrality measure which account for the temporal-topological characteristics introduced by the shortest time-respecting path structures in real-world temporal networks.
Comparing the importance of nodes according to i) temporal centralities, ii) centralities calculated based on a commonly used static, time-aggregated representation, and iii) second-order centralities calculated based on a static, second-order time-aggregated representation, we show that higher-order aggregate networks provide interesting perspectives for the analysis of temporal networks.
We finally conclude our article by a summary of key contributions and a discussion of open issues and future work.

\newpage

\section{Temporal Networks and Time-respecting Paths}
\label{sec:definitions}

In this  section, we formally introduce the basic concepts and definitions used throughout our work.
In particular, we define the notion of a \emph{temporal network} used throughout this article, as well as \emph{time-respecting paths} which are the basis for the notions of \emph{distances} and \emph{path-based centralities} in temporal networks which will be used in subsequent sections.

\subsection{Temporal, time-aggregated and time-unfolded networks}
We define a temporal network $G^T=(V,E^T)$ as a tuple consisting of a set of nodes $V$ and a set $E^T \subseteq V \times V \times [0,T]$ of time-stamped links $(v,w;t) \in E^T$ for an observation period $[0,T]$.
Importantly, we assume \emph{discrete} time stamps $t \in [0,T]$ and time-stamped links $(v,w;t)$ which indicate the presence of the link $(v,w)$ at time $t$.
This ``instantaneous'' definition particularly does not allow links to be assigned a \emph{duration}, i.e. we cannot directly assign links a time interval during which they exist.
However, we can nevertheless represent links that persist for some time interval $\left[t_{start},t_{end}\right]$ by assuming some small unit of discrete time $\Delta t$ and adding multiple time-stamped links $(v,w;t)$ at time stamps $t=t_{start}, t_{start}+\Delta t, t_{start}+2 \Delta t, \ldots, t_{end}$.
These assumptions naturally lend themselves to real-world time-stamped data sets, which are typically obtained based on some sort of \emph{sampling}, whose sampling frequency defines the smallest unit of time $\Delta t$.
\begin{figure}
    \subfigure[Temporal network $G_1$\label{fig:example:G1}]{\includegraphics[width=.33\textwidth]{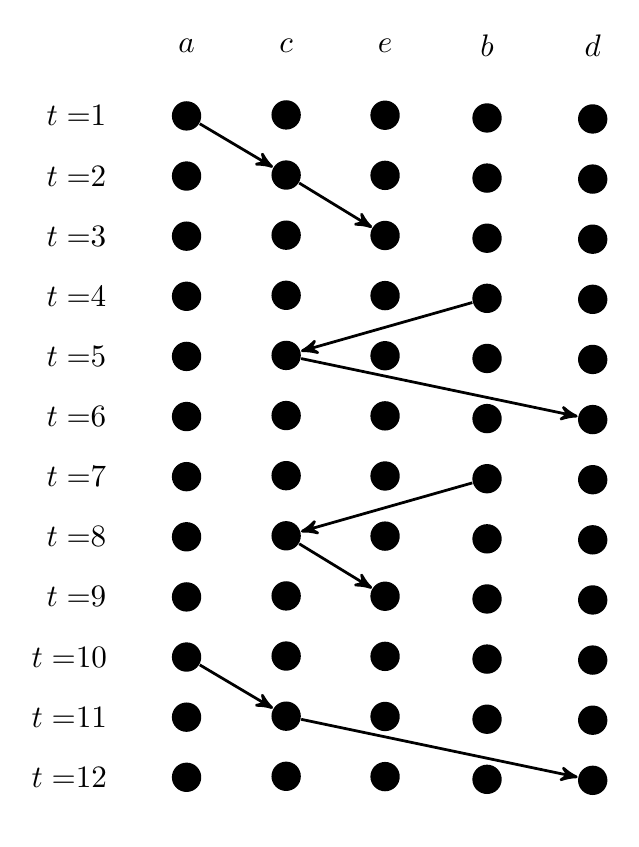}}
    \subfigure[Temporal network $G_2$\label{fig:example:G2}]{\includegraphics[width=.33\textwidth]{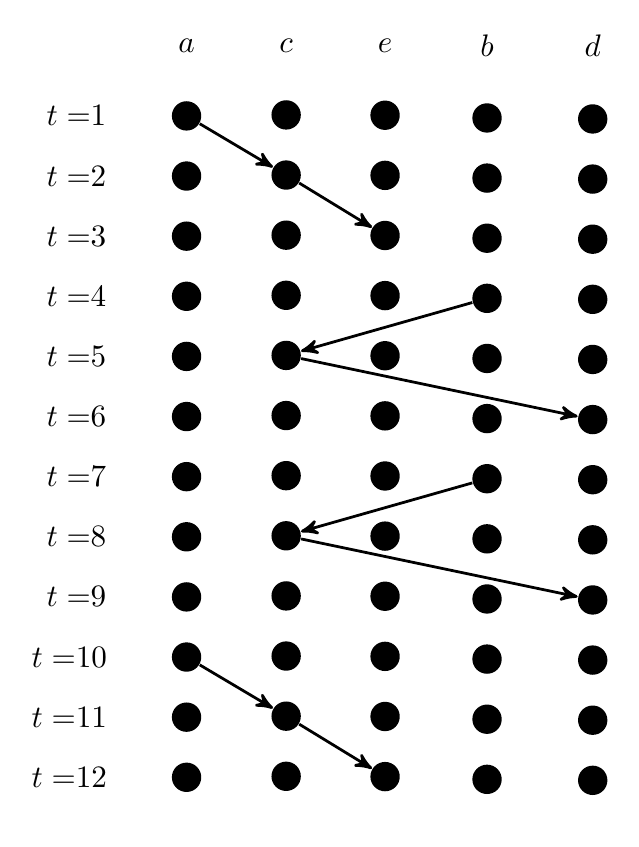}}
    \subfigure[Weighted, time-aggregated representation of both $G_1$ and $G_2$\label{fig:example:agg}]{\includegraphics[width=.3\textwidth]{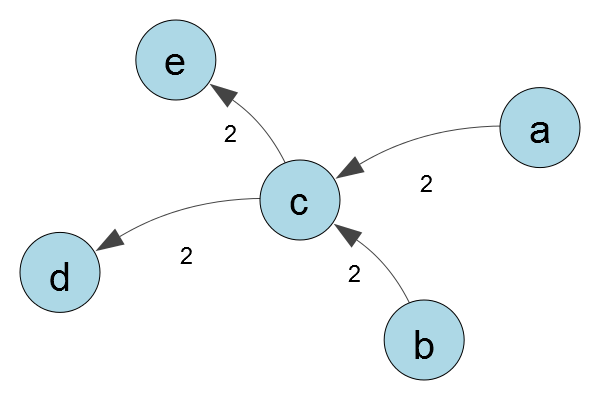}}
    \caption{Time-unfolded and weighted static, time-aggregated representation of two temporal networks $G_1$ and $G_2$}
    \label{fig:example}
\end{figure}

For illustrative purposes it is often useful to be able to visualize temporal networks.
Throughout this article, we will use so-called \emph{time-unfolded networks}, a simple and intuitive static representation of temporal networks which, in different variants, has been used in a number of previous works~\cite{Kim2012,Pfitzner2013,Scholtes2014,Takagutchi2015}.
The key idea of this two-dimensional static representation is to arrange all nodes on a horizontal dimension, while unfolding time to an additional vertical dimension as illustrated in Fig.~\ref{fig:example}.
For an observation period $\left[0, \ldots, T\right]$ and a given $\Delta t$ we can then add \emph{temporal copies} of all nodes for all possible time steps $k \Delta t$ (for $k=0, 1, \ldots$).
For simplicity, in the following we assume $\Delta t = 1$, which allows us to denote the temporal copies of a node $v$ as $v_t, v_{t+1}, v_{t+2}, \ldots$.
The main benefit of this construction is that it allows us to represent a time-stamped link $(v,w;t)$ by means of a static link $(v_t,w_{t+1})$ connecting the temporal copies $v_t$ and $w_{t+1}$ of node $v$ and node $w$ respectively.
The intuition behind this notation is that a quantity residing at node $v$ at time $t$ can move to node $w$ via a time-stamped link $(v,w;t)$, arriving there at the next time step $t+1$.
Two simple examples for time-unfolded static representations of two different temporal networks with five nodes and eight time-stamped links are shown in Fig.~\ref{fig:example:G1} and \ref{fig:example:G2}.

Despite the recent development of methods to study temporal networks, the most wide-spread way to study time-stamped network data is to aggregate all time-stamped links into a static, \emph{time-aggregated network} $G=(V,E)$.
This means that, given a temporal network $G^T=(V,E^T)$, two nodes $v,w \in V$ are connected in the static network whenever a time-stamped link exists at \emph{any} time stamp, i.e., $(v,w) \in E$ iff $(v,w;t) \in E^T$ for any $t \in \left[0, T\right]$.
Additional information about the statistics of time-stamped links in the underlying temporal network can be preserved by considering a \emph{weighted time-aggregated} network, in which weights $\omega(v,w)$  indicate the number of times time-stamped links $(v,w;t)$ have been active during the observation period.
I.e., we consider a weighted time-aggregated network with a weight function $\omega:E \rightarrow \mathbb{N}$ defined as
\begin{align*}
  \omega(v,w) := |\{ t \in \left[0,T\right] | (v,w;t) \in E^T \}|.
\end{align*}
Figure~\ref{fig:example:agg} shows the weighted, time-aggregated networks corresponding to the two temporal networks shown in Fig.~\ref{fig:example:G1} and \ref{fig:example:G2}.
These simple examples highlight the important fact that \emph{different} temporal networks are consistent with the \emph{same} weighted, time-aggregated network.
This is due to the fact that in the time-aggregated network we lose all information on both the timing and the ordering of links in the temporal network.

\subsection{Time-respecting paths}
Importantly, both the timing and the ordering of links influence path structures in temporal networks.
In particular, in the context of temporal networks we must consider time-respecting paths, an extension of the concept of paths in static network topologies which additionally respects the timing and ordering of time-stamped links~\cite{kempe2000connectivity,Pan2011,Holme2012}.
For the remainder of this paper, we define a \emph{time-respecting path} between a source node $v$ and a target node $w$ to be any sequence of time-stamped links
\begin{align*}
  (v_0,v_1; t_1), (v_1,v_2; t_2)\ldots, (v_{l-1}, v_l; t_l)
\end{align*}
such that $v_0=v, v_l=w$ and the sequence of time-stamps is increasing, i.e. $t_1 < t_2 \ldots < t_l$.
The latter condition on the ordering of links is particularly important since it is a necessary condition for causality.
This means that for any temporal network a node $a$ is able to influence node $c$ based on two time-stamped links $(a,b)$ and $(b,c)$ only if link $(a,b)$ has occurred \emph{before}  link $(b,c)$.
A simple example for a time-respecting path $(a,c;1),\ (c,d;4)$ can be seen in Fig.~\ref{fig:example:G1}, where the time-unfolded representation of the temporal network $G_1$ is illustrated.

At this point, it is important to note that, different from the usual notion of paths in static networks, the question whether a time-respecting path exists between two nodes requires to specify a \emph{start time} $t_0 \leq t_1$.
In the example of Fig.~\ref{fig:example:G1} we observe a time-respecting path $(a,c;t_1=1), (c,d;t_2=5)$ between node $a$ and $d$, which can only be taken if we consider paths starting at node $a$ at time $t_0=1$.
If instead we were to ask for a time-respecting path between $a$ and $d$ starting at node $a$ at time $t_0=5$, our only choice would be the path $(a,c;10), (c,d;11)$.

\subsection{Time-respecting paths with a maximum time difference}
In the definition of a time-respecting path above, we have required that the sequence of time stamps of the links constituting the path must be increasing.
Clearly, this condition is rather weak since it makes no assumptions whatsoever about the time difference between two consecutive time-stamped links on a time-respecting path.
As such, for the mere existence of a time-respecting path in a temporal network evolving over a period of years, it is actually not important whether the time difference between two consecutive links is a few seconds or a few years.

However, we typically study time-respecting path structures because they constitute the substrate for the evolution of dynamical processes which have intrinsic time scales that are much smaller than the period during which we observe a temporal network.
In the study of time-respecting paths, it is thus often reasonable to impose a \emph{maximum time difference} $\delta$, i.e. we limit the temporal gaps between two consecutive time-stamped links that are considered to contribute to a time-respecting path to a maximum of $\delta$~\cite{Pan2011,Holme2012}.
In this case, rather than requiring a mere increasing sequence of time-stamps, we demand that the condition $0 < t_{i+1} - t_i \leq \delta$ must be fulfilled for all $i=1, \ldots, l-1$.
For a maximum time difference of $\delta=1$, we thus limit ourselves to the study of time-respecting paths for which all time-stamped links occur at immediately consecutive time stamps.
As another limiting case, we can consider $\delta=\infty$, which means that we impose no further condition apart from the requirement the the sequence of time stamps of links on a time-respecting path is increasing
Revisiting the example of Fig.~\ref{fig:example:G1}, we observe that the time-respecting path $(a,c;1), (c,d;5)$ only exists if we allow for a maximum time difference $\delta=4$, while for all $\delta<4$ the only time-respecting path between the nodes $a$ and $d$ is $(a,c;10), (c,d;11)$.

\subsection{Shortest and fastest time-respecting paths}
Let us now formally define the length of time-respecting paths in a temporal network, which will allow us to define the notion of \emph{shortest time-respecting} paths used throughout our work.
Due to the additional temporal dimension, the length of a time-respecting path
\begin{align*}
  (v_0,v_1;t_1),\ldots,(v_{l-1},v_l;t_l)
\end{align*}
can be studied both from a topological and a temporal perspective.
Following the usual terminology, we call the number $l$ of time-stamped links on a time-respecting path the (topological) \emph{length} of the path.
We further call the time difference $t_l-t_1+1$ the \emph{duration} of the path.
Here the increment by one accounts for the duration of the final link $(v_{l-1},v_{l};t_l)$, i.e. for the fact that any process starting at node $v_0$ at time $t_1$ will only reach node $v_l$ at time $t_{l+1}$.

Having defined both the length and duration of time-respecting paths, it is now trivial to define the \emph{shortest time-respecting path} between two nodes $v$ and $w$ as the time-respecting path with the smallest (topological) length.
In analogy, we define the \emph{fastest time-respecting path} as the time-respecting path with the smallest (temporal) duration.
Following our previous comment about the necessity to define a start time $t_0$ for a time-respecting path, it is clear that the shortest or fastest time-respecting path can only be found unambiguously with respect to a given start time $t_0$, i.e. at different times during the evolution of a temporal network the same pair of nodes can be connected by different shortest or fastest time-respecting paths.

\subsection{Transitivity of paths in static and temporal networks}
Let us conclude this preliminary section by highlighting important differences between paths in static networks compared to time-respecting paths in temporal networks, that result from the ordering and timing of links.
Let us first highlight that paths in static networks are \emph{transitive}.
This means that from the presence of two paths $(v_0, v_1), \ldots, (v_{k-1}, v_k)$ and $(v_k, v_{k+1}), \ldots, (v_{l-1}, v_l)$ between $v_0$ and $v_k$ and between $v_k$ and $v_l$ respectively, we can conclude that a path $(v_0, v_1), \ldots, (v_{l-1}, v_{v_l})$ between nodes $v_0$ and $v_l$ necessarily exists\footnote{Note though that this transitive path may or may not be the shortest path between the two nodes.}.
This transitivity has the important mathematical consequence that the entries in the $k$-th power $A^k$ of the adjacency matrix $A$ of a static network topology count all possible paths of length $k$ between all possible pairs of nodes.
Furthermore, transitivity of paths is the basis for a wealth of \emph{algebraic network-analytic methods} such as spectral partitioning, the analysis of dynamical processes based on eigenvectors and eigenvalues, or the computation of centrality measures that are based on eigenvalue problems.

Notably, the property of transitivity of paths in static networks does \emph{not} extend to time-respecting paths in temporal networks.
Here, two time-respecting paths \\ $(v_0,v_1;t_1), \ldots, (v_{k-1}, v_k; t_k)$ and \\ $(v_k,v_{k+1};t_{k+1}), \ldots, (v_{l-1}, v_l; t_l)$ only translate into a time-respecting path between $v_0$ and $v_l$ if $t_k < t_{k+1}$ and, assuming that we impose a maximum time difference $\delta$, if $0 < t_{k+1}-t_k \leq \delta$.

The simple observation that transitivity of paths holds in static networks, while it does not necessarily hold in temporal networks implies that by an analysis of static, time-aggregated networks, we may overestimate transitivity in temporal networks.
We can again illustrate this using our simple example of Fig.~\ref{fig:example}, which shows two temporal networks $G_1$ and $G_2$ that are both consistent with the same (weighted) time-aggregated network shown in~Fig.~\ref{fig:example:agg}.
Here, judging from the presence of a path $(a,c),(c,d)$ in the time-aggregated network, we may think that a time-respecting path connecting node $a$ to $d$ exists in the underlying temporal network.
Looking at the two temporal networks $G_1$ and $G_2$ shown in Fig.~\ref{fig:example:G1} and Fig.~\ref{fig:example:G2} respectively, we see that at least for small values for the maximum time difference $\delta$ (such as $\delta=1$) a corresponding time-respecting path only exists in the temporal network $G_1$, while it is absent in $G_2$.

\section{Higher-Order Aggregate Networks}
\label{sec:higherOrder}

In the previous section we have seen that for large maximum time differences $\delta$ we expect the shortest time-respecting paths to be rather similar to the shortest path in a static, time-aggregated representation.
This is an intuitive result since by using large maximum time differences $\delta$, we apply an implicit ``aggregation'' of time stamps which may nevertheless be far apart in the temporal dimension.
At the same time, we observe that for small values of $\delta$ the temporal characteristics of the network result in time-respecting path structures that are markedly different from those in the static, time-aggregated network.
As argued above, this implies that dynamical processes which evolve at time scales similar to that of the temporal network will be significantly affected by these path structures.
It further questions the usefulness of path-based centrality measures that are computed based on the commonly used time-aggregated representation of temporal networks.

In this section, we introduce \emph{higher-order time-aggregated networks}, a simple yet powerful abstraction of temporal networks which can be used to address some of the aforementioned problems.
It can be seen as a simple generalization of the usual \emph{first-order} time-aggregated representation introduced in Section~\ref{sec:definitions}, and it has recently been shown to provide interesting insights about the evolution of dynamical processes in temporal networks~\cite{Scholtes2014}.

\subsection{$k$-th order aggregate networks}
The key idea behind this abstraction is that the commonly used time-aggregated network is the simplest possible time-aggregated representation whose weighted links captures the frequencies of time-stamped links.
Considering that each time-stamped link is a time-respecting path of length one, it is easy to generalize this abstraction to \emph{higher-order time-aggregate networks} in which weighted links capture the frequencies of longer time-respecting paths.
For a temporal network $G^T=(V,E^T)$ we thus formally define a $k$-th order time-aggregated (or simply aggregate) network as a tuple $G^{(k)} = (V^{(k)}, E^{(k)})$ where $V^{(k)} \subseteq V^k$ is a set of node $k$-tuples and $E^{(k)} \subseteq V^{(k)} \times V^{(k)}$ is a set of links.
For simplicity, we call each of the $k$-tuples $v=v_1-v_2-\ldots-v_k$ ($v \in V^{(k)}, v_i \in V$) a \emph{$k$-th order node}, while each link $e \in E^{(k)}$ is called a \emph{$k$-th order link}.
We further assume that a $k$-th order link $(v,w)$ between two $k$-th order nodes $v=v_1-v_2-\ldots-v_k$ and $w=w_1-w_2-\ldots-w_k$ exists if they overlap in exactly $k-1$ elements such that $v_{i+1}=w_i$ for $i=1, \ldots, k-1$.
The basic idea behind this construction is that each $k$-th order link $(v,w)$ represents a possible time-respecting path of length $k$ in the underlying temporal network, which connects node $v_1$ to node $w_k$ via $k$ time-stamped links
\begin{align}
  (v_1,v_2=w_1;t_1), \ldots, (v_{k}=w_{k-1}, w_k; t_k)
  \label{eq:pathsK}
\end{align}
In analogy to the weights in a usual (first-order) aggregate representation, we further define the weights of such $k$-th order links by the frequency of the underlying time-respecting paths in the temporal network.
Considering a maximum time difference $\delta$ and two $k$-th order nodes $v=v_1-v_2-\ldots-v_k$ and $w=w_1-w_2-\ldots-w_k$ we thus define
\begin{align*}
  \omega(v,w) := | P(v,w,\delta) |
\end{align*}
where
\begin{align*}
  \begin{split}
  P = &\{ (v_1,v_2=w_1;t_1), \ldots \\ & \ldots, (v_{k}=w_{k-1}, w_k; t_k) : 0 < t_{i+1}-t_i \leq \delta \}
  \end{split}
\end{align*}
is the set of all time-respecting paths in the temporal network that i) consist of the sequence of links indicated in Eq.~\ref{eq:pathsK}, and ii) are consistent with a given maximum time difference of $\delta$.

The higher-order aggregate network construction introduced above has a number of advantages.
First and foremost, it provides a simple static abstraction of a temporal network which can be studied by means of standard methods from (static) network analysis.
Each static path of length $l$ in a $k$-th order aggregate network can be mapped to a time-respecting path of length $k+l-1$ in the original network.
Importantly, and different from a first-order representation, $k$-th order aggregate networks allow to capture \emph{non-Markovian characteristics} of temporal networks.
In particular, they allow to represent temporal networks in which the $k$-th time-stamped link $(v_{k}=w_{k-1}, w_k)$ on a time-respecting path depends on the $k-1$ previous time-stamped links on this path.
With this, we obtain a simple static network topology that contains information both on the presence of time-stamped links in the underlying temporal network, as well as on the \emph{ordering} in which sequences of $k$ of these time-stamped links occur.

\subsection{Example: second-order aggregate networks}
In the following, we illustrate our approach by constructing second-order aggregate representations of the two temporal networks $G_1$ and $G_2$ shown in Fig.~\ref{fig:example}.
Both $G_1$ and $G_2$ are consistent with the same first-order time-aggregated network.
We can easily generate second-order \\ time-aggregated networks of the two temporal networks by extracting all time-respecting paths of length two (and assuming a given maximum time difference $\delta$).
For simplicity, in the following we limit our study to $\delta=1$.
For the temporal network $G_1$ shown in Fig.~\ref{fig:example:G1}, we observe the following four different time-respecting paths of length two:
\begin{align*}
  (a,c;1), (c,e;2) \\
  (b,c;3), (c,d;4) \\
  (b,c;7), (c,e;8) \\
  (a,c;10), (c,d;11)
\end{align*}
Based on the definition of links and link weights outlined above, we thus obtain the following four weighted second-order links:
\begin{align*}
  \omega(a-c, c-e)=1 \\
  \omega(b-c, c-d)=1 \\
  \omega(b-c, c-e)=1 \\
  \omega(a-c, c-d)=1
\end{align*}
The resulting second-order network is depicted in Fig.~\ref{fig:example2nd:G1}
Applying the same methodology to the temporal network $G_2$ shown in Fig.~\ref{fig:example:G2} we obtain the following four time-respecting paths of length two
\begin{align*}
  (a,c;1), (c,e;2) \\
  (b,c;3), (c,d;4) \\
  (b,c;7), (c,d;8) \\
  (a,c;10), (c,e;11) \\
\end{align*}
from which we obtain the following two weighted second-order links:
\begin{align*}
  \omega(a-c, c-e)=2 \\
  \omega(b-c, c-d)=2
\end{align*}
The resulting second-order aggregate network is shown in Fig.~\ref{fig:example2nd:G2}.
\begin{figure}[htb]
  \centering
  \subfigure[Temporal network $G_1$ \label{fig:example2nd:G1}]{\includegraphics[width=.4\textwidth]{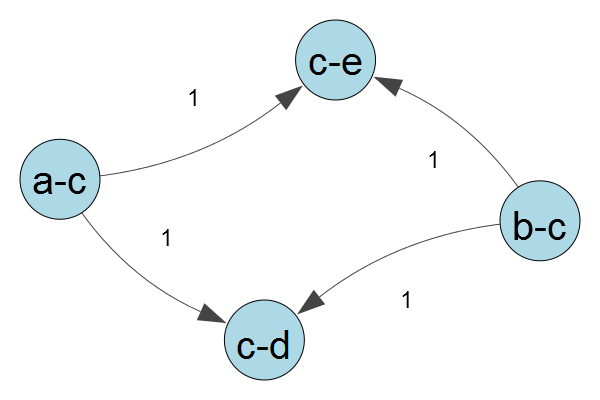}} \qquad
  \subfigure[Temporal network $G_2$\label{fig:example2nd:G2}]{\includegraphics[width=.4\textwidth]{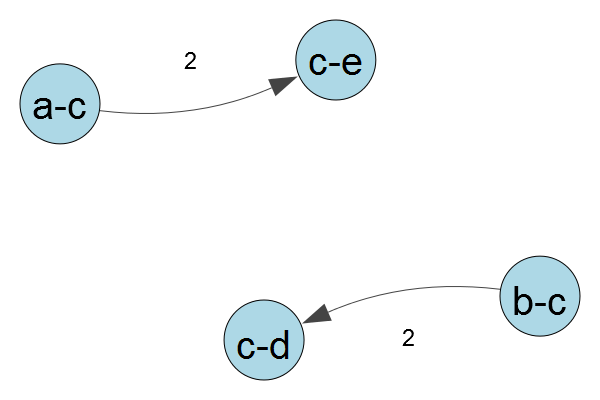}}
  \caption{Second-order aggregate networks $G^{(2)}$ corresponding to the two temporal networks shown in Fig.~\ref{fig:example}.}
\end{figure}
Here we observe that even though the two temporal networks $G_1$ and $G_2$ only differ in the order of two time-stamped links, the resulting second-order aggregate network is markedly different.
The second-order network of $G_1$ indicates time-respecting paths connecting node $a$ to both nodes $e$ and $d$ (both paths passing via node $c$).
In particular, this corresponds to the connectivity that we would expect based on the transitivity of static paths in the first-order aggregate network shown in Fig.~\ref{fig:example:agg}.
The second-order network shown in Fig.~\ref{fig:example2nd:G2} reveals that the transitive path $(a,c),(c,d)$ in the first-order aggregate network does not translate to a time-respecting path in the temporal network $G_2$.

Clearly, the second-order aggregate networks illustrated above are only a special, particularly simple type of general, higher-order aggregate networks.
Nevertheless, in the following section we will demonstrate that it contains important information about the causal topology of temporal networks which can help us in the analysis of temporal networks.
%

In what follows, we will thus provide an in-depth study of second-order aggregate representations of six empirical data sets that will be introduced in the following section.
Here, we will particularly focus on the question how second-order aggregate networks can foster the calculation of approximate measures for path-based node centralities in temporal networks.


\section{Temporal Node Centralities in Second-Order Aggregate Networks}
\label{sec:centrality}

Having introduced the abstraction of higher-order aggregate networks in section~\ref{sec:higherOrder}, let us now demonstrate the use of a \emph{second-order aggregate representation} for the study of path-based centralities in temporal networks.
We will study this question using the following six, publicly available empirical data sets representing different types of temporal networks:
(AN) covers time-stamped antenna-antenna interactions inferred from a filming of ants in an ant colony~\cite{Blonder2011};
(EM) represents time-stamped E-Mail exchanges between employees in a manufacturing company~\cite{Michalski2011};
(HO) covers time-stamped proximity interactions between patients and medical staff in a hospital~\cite{Vanhems2013};
(RM) is based on time-stamped social interactions between students and academic staff at a university campus~\cite{Eagle2006};
(LT) has been reconstructed from data on passenger itineraries in the London Tube metro system available through the Rolling Origin and Destination Survey of the Transport of London~\cite{LTdata}, and
(FL) was constructed based from data on flight itineraries of passengers on domestic flights in the United States available from the Bureau of Transportation Statistics~\cite{FLdata}.
A detailed description about the processing of these data sets and the extraction of time-stamped network data is available in~\cite{Scholtes2014}, which is why we omit an elaborate discussion here.

Regarding the choice of a reasonable maximum time difference $\delta$ for the notion of shortest time-respecting paths as discussed in section~\ref{sec:definitions}, we emphasize that the choice of  this parameter needs to be adapted to the inherent time scale of the network evolution in each of the six data sets individually.
In general, such a choice is non-trivial as it heavily influences i) whether or not pairs of nodes can reach each other, and ii) to what extent temporal characteristics influence the structures of time-respecting paths.
In particular, for too small choices of $\delta$ the definition of time-respecting paths is likely to be too restrictive and almost no paths will be found~\cite{Pan2011,Holme2012}.
Contrariwise, the choice of a too large value for $\delta$ results in the fact that we effectively ``aggregate'' the time-stamped sequence of links, thus discarding information about the detailed ordering and timing of links.
For our analysis, for each of the six data sets individually, we have thus chosen the minimum parameter $\delta$ for which we still obtain a topology of time-respecting paths that is strongly connected, thus ensuring that we can compute reasonable measures of path-based centralities while retaining as much of the temporal characteristics as possible (c.f. details in \cite{Scholtes2014}).

In the remainder of this section, we will focus our analysis on three widely adopted path-based notions of centrality, namely i) betweenness, ii) closeness and iii) reach centrality.
The rationale behind this choice is that all of these three measures can easily be computed based on paths in time-aggregated networks, while they additionally facilitate a straight-forward extension to temporal networks based on the notion of shortest time-respecting paths (c.f. similar extensions studied in ~\cite{Kim2012,Holme2012,Takagutchi2015}).
In the following, we first formally define the \emph{temporal} betweenness, closeness and reach centrality of nodes.
We then compute the resulting measures for all nodes based on the actual shortest time-respecting paths in the time-stamped link sequences in our six data sets (and using the individually determined maximum time difference $\delta$).
The resulting centrality scores are considered as the \emph{ground-truth} against which we then compare the centrality scores resulting from the application of the same centrality measures to i) the commonly used (first-order) time-aggregated representation, and ii) a second-order aggregate network representation of the corresponding temporal network.


\subsection{Temporal Betweenness Centrality}

We first address the question to what extent the temporal betweenness centrality of nodes in a temporal network can be approximated by means of static betweenness centralities calculated based on static, time-aggregated representations.
To this end, we first formally define the temporal betweenness centrality of a node in a temporal network.
According to the common definition, the (unnormalized) betweenness centrality of a node $v$ is simply calculated as the total number of shortest paths passing through node $v$~\cite{Freeman1977}.
Highlighting the fact that we can directly apply this measure to first-order time-aggregated networks, we thus define the \emph{first-order betweenness centrality} $\text{BC}^{\text{(1)}}(v)$ of a node $v$ as
\begin{align}
    \text{BC}^{\text{(1)}}(v) := \sum_{u\neq v \neq w} |P^{(1)}(u,w;v)|
    \label{eq:betweenness:1st}
\end{align}
where $P^{(1)}(u,w;v)$ denotes the set of those shortest paths from node $u$ to $w$ in a static network that pass through node $v$.

Applying this idea to temporal networks, a straight-forward way to define the \emph{temporal betweenness centrality} of a node is to count all shortest \emph{time-respecting paths} passing through it.
However, and as mentioned in Section~\ref{sec:definitions}, temporal networks introduce the complication that, in order to unambiguously define shortest time-respecting paths, we need to include a start time $t_0$ starting from which time-respecting paths are to be considered.
For each pair of nodes $u,w$ and each start time $t_0$ we can thus directly define an instantaneous distance function for a temporal network as
\begin{align}
    \text{dist}^{\text{temp}}(u,v,t_0) := \text{len}(p), p \in P^{\text{temp}}(u,v,t_0)
    \label{eq:distance:tempInst}
\end{align}
where $P^{\text{temp}}(u,v,t_0)$ denotes the set of shortest \\ time-respecting paths from $u$ to $v$ that start at time $t_0$ (and which are consistent with a given maximum time difference $\delta$).
Based on this instantaneous definition of shortest time-respecting paths, we can further define a distance function that gives the minimum distance across \emph{any} start time as follows:
\begin{align}
  \text{dist}^{\text{temp}}(u,v) := \min_{t_0} \text{dist}^{\text{temp}}(u,v,t_0)
    \label{eq:distance:temp}
\end{align}
With this we can further define the set of shortest time-respecting paths across all start times as
\begin{align}
\scriptsize
  P^{\text{temp}}(u,v) := \bigcup_{t_0} \{ p \in P^{\text{temp}}(u,v,t_0) | \text{len}(p) = \text{dist}^{\text{temp}}(u,v) \}
\end{align}
i.e. we only consider those (instantaneous) shortest time-respecting paths whose lengths correspond to the minimum shortest time-respecting length across \emph{all} possible start times.
We can now define the \emph{temporal betweenness centrality} $\text{BC}^{\text{temp}}(v)$ of a node $v$ in analogy to Eq.~\ref{eq:betweenness:1st} as
\begin{align}
    \text{BC}^{\text{temp}}(v) := \sum_{u\neq v \neq w} |P^{\text{temp}}(u,w;v)|
\end{align}
where $P^{\text{temp}}(u,w;v)$ denotes the set of those shortest time-respecting paths across all start times which connect node $u$ to $w$ and which pass through node $v$.

Let us illustrate this definition using the temporal networks shown in Fig.~\ref{fig:example:G1} and Fig.~\ref{fig:example:G2}.
Applying the static betweenness centrality as defined in Eq.~\ref{eq:betweenness:1st} to the first-order aggregate network shown in Fig.~\ref{fig:example:agg}, we find that for node $c$ we have $\text{BC}^{\text{(1)}}(c)=4$, while for all other nodes we have a betweenness centrality of zero.
Again assuming $\delta=1$, for the temporal betweenness centrality of node $c$ in network $G_1$ shown in Fig.~\ref{fig:example:G1}, we find that indeed four shortest time-respecting paths pass through node $c$, i.e. we have $\text{BC}^{\text{temp}}(c)=4$ while we again have a zero temporal betweenness centrality for all other nodes.
Notably, in this particular case the temporal betweenness centralities of nodes correspond to the betweenness centralities of nodes calculated based on the first-order time-aggregated network.
This happens because all paths in the first-order aggregate network have a counterpart in terms of a shortest time-respecting path.

However, in section~\ref{sec:definitions} we have seen that, in general, shortest time-respecting paths in temporal networks may not coincide with shortest paths in the (first-order) time-aggregated network.
As a consequence, the temporal betweenness centralities of nodes may differ from the first-order betweenness centralities calculated from a static, first-order aggregate representation.
This can be seen for the temporal network $G_2$ shown in Fig.~\ref{fig:example:G2}.
Based on the temporal sequence of time-stamped links, here we find only two different shortest time-respecting paths passing through node $c$, namely one connecting node $a$ via $c$ to $e$ and a second one connecting node $b$ via $c$ to $d$.
The two additional shortest time-respecting paths found in $G_1$ are absent in $G_2$, therefore in $G_2$ node $c$ has a temporal betweenness centrality $\text{BC}^{\text{temp}}(c)=2$, thus being, at least from the perspective of temporal betweenness centrality, less important than in $G_1$.

In the following we study the question to what extent first-order betweenness centralities can be used as a proxy for the temporal betweenness centralities of nodes in our six data sets of real-world temporal networks.
In particular, we study this question in the following way:
For each node $v$ in the six data sets we calculate i) the first-order betweenness centrality $\text{BC}^{(1)}(v)$ based on the first-order aggregate network, as well as ii) the (ground truth) temporal betweenness centrality $\text{BC}^{\text{temp}}(v)$ based on actual shortest time-respecting paths in the temporal network.
We then assess the correlation between both measures by computing the Pearson correlation coefficient (as well as the corresponding p-value) for the sequence of paired values $(\text{BC}^{(1)}(i), \text{BC}^{\text{temp}}(i))$ for all nodes $i \in V$.

Since centrality scores of nodes in networks are often used and interpreted in a relative fashion, we further perform an additional analysis that accounts for variations in the actual centrality values, which however may not affect the relative importance of nodes.
For this, we first rank nodes according to their temporal and first-order betweenness centralities respectively.
We then calculate the Kendall-Tau rank correlation coefficient in order to quantitatively assess to what extent nodes are ranked similarly according to both notions of centrality (even though the actual centrality values for these nodes may differ).

The results of this analysis are shown in the left column of Table~\ref{tab:betweenness}, in which we report both the Pearson as well as the Kendall-Tau rank correlation coefficients between the temporal and the first-order betweenness centralities of nodes for each of the six data sets introduced above.
Here, a first interesting result is that both the Pearson and the Kendall-Tau rank correlation coefficients exhibit a large variation between $0.75$ and $0.99$, as well as $0.59$ and $0.81$ respectively.
\begin{table*}[t]
\centering
\begin{tabular}{|l|l|l|l|l|}
  \hline
  {}            &  \multicolumn{2}{|c|}{$\text{BC}^{\text{temp}} \thicksim \text{BC}^{(1)}$} &  \multicolumn{2}{|c|}{$\text{BC}^{\text{temp}} \thicksim \text{BC}^{(2)}$} \\
  {}                & Pearson           & Kendall-Tau       & Pearson           & Kendall-Tau \\
  \hline
  E-Mail (EM)       & 0.80 (3.29e-22)   & 0.73 (8.36e-26)   & 0.97 (7.52e-60)   & 0.74 (1.11e-26)\\
  Ants (AN)         & 0.82 (3.49e-16)   & 0.64 (2.05e-13)   & 0.80 (1.96e-14)   & 0.59 (1.94e-11)\\
  Hospital (HO)     & 0.93 (2.39e-23)   & 0.81 (1.18e-17)   & 0.96 (2.36e-30)   & 0.87 (5.55e-20)\\
  RealityMining (RM)& 0.95 (2.83e-30)   & 0.62 (7.28e-12)   & 0.93 (3.74e-26)   & 0.75 (1.12e-16)\\
  London Tube (LT)  & 0.85 (2.58e-37)   & 0.66 (1.22e-29)   & 0.87 (3.28e-42)   & 0.71 (9.32e-34)\\
  Flights (FL)      & 0.99 (6.91e-108)  & 0.66 (9.09e-26)   & 0.99 (2.66e-98)   & 0.65 (4.25e-25)\\
  \hline
\end{tabular}
\caption{Pearson and Kendall-Tau rank correlation coefficients between temporal betweenness centrality (ground truth) and betweenness centrality calculated based on the first-order aggregate network and the second-order aggregate network. Values in parentheses indicate the p-value.}
\label{tab:betweenness}
\end{table*}
The results indicate that, depending on the characteristics of the underlying temporal network, temporal betweenness centralities can be reasonably well approximated by first-order betweenness centrality for some data sets (e.g., for (FL), (HO), (RM)) while such an approximation should be taken with caution for other data sets.

Based on these results it is reasonable to ask if we can better approximate temporal centrality, especially for those data sets where the correlation between the first-order and the temporal betweenness centrality is comparably weak.
In Section~\ref{sec:higherOrder} we have argued that the generalization of higher-order aggregate networks allows to construct static representations of temporal networks that capture both temporal and topological characteristics that emerge from the ordering of links and the statistics of time-respecting paths.
Focusing on a second-order representation, in the remainder of this section we will study to what extent second-order aggregate networks can be used in the analysis of temporal node centralities.

Importantly, such an analysis is facilitated by the fact that second-order aggregate networks are \emph{static networks}, which allows for a straight-forward application of standard centrality measures to the second-order topology.
In the case of second-order aggregate networks, applying standard centrality measures we obtain centrality values for higher-order nodes $(v,w)$, each of the higher-order nodes being a $k$-tuple of nodes in the first-order network.
In order to arrive at a centrality measure for the original (first-order) nodes, we thus must project this measure to the level of nodes in the first-order network.

Luckily, this can be done in a simple way which we outline in the following:
For a second-order network $G^{(2)}=\left(V^{(2)}, E^{(2)}\right)$, let us first define a second-order distance function $\text{dist}^{(2)}(v,w)$ which, for each pair of \emph{first-order} nodes $v,w \in V^{(1)}$, gives the length of a shortest path based on the topology of the \emph{second-order} aggregate network as
\begin{align}
  \text{dist}^{(2)}(v,w) := \min_{\substack{x,y \in V^{(2)}\\x=v-*\ \ \\ y=*-w\ \ }} L^{(2)}(x,y)+1
  \label{eq:distance:2nd}
\end{align}
where $L^{(2)}(x,y)$ denotes the length of a shortest path between the second-order nodes $x,y \in V^{(2)}$.
The rationale behind this definition is that in the second-order aggregate network, we can have multiple shortest paths with different lengths between different second-order nodes, which nevertheless map to paths between a single pair of first-order nodes.
As an example, consider the two first-order nodes $a$ and $d$ in the simple second-order network shown in Fig.~\ref{fig:exampleG2}.
\begin{figure}
\centering
  \includegraphics[width=.5\textwidth]{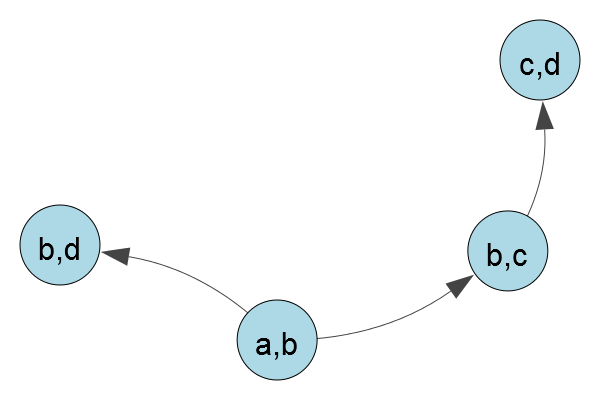}
  \caption{Simple example for a second-order aggregate network}
  \label{fig:exampleG2}
\end{figure}
Here we observe that, from the perspective of second-order nodes, both $(a-b, b-d)$ as well as \\ $(a-b, b-c), (b-c, c-d)$ are shortest paths (between different pairs of nodes) in the second-order network with lengths $L^{(2)}(a-b, b-d) = 1$ and \\ $L^{(2)}(a-b, c-d) = 2$ respectively.
However, from the perspective of first-order nodes both of these second-order paths connect node $a$ to node $d$ (via paths of length $2$ and $3$ respectively).
Using the definition from Eq.~\ref{eq:distance:2nd} thus allows us to correctly calculate the second-order distance between $a$ and $d$ as $\text{dist}^{(2)}(a,d)= L^{(2)}(a-b, b-d) +1 = 2$.

The above definition of a second-order distance function now allows us to define a \emph{second-order betweenness centrality} $\text{BC}^{(2)}(v)$ of a node $v$ based on Eq.~\ref{eq:betweenness:1st}.
For this, we simply count all second-order shortest paths between two nodes $u$ and $w$ which i) pass through node $v$, and ii) whose length corresponds to the second-order distance $\text{dist}^{(2)}(u,v)$.
Formally, we define
\begin{align}
    \begin{split}
       & \text{BC}^{(2)}(v) :=  \\ & \sum_{\substack{x\neq y \in V \\ u-x \in V^{(2)}\\ y-w \in V^{(2)}}} \hspace{-0.3cm} |\{ p \in P^{(2)}(u-x,y-w;v):  \text{len}(p) =\text{dist}^{(2)}(u,w) \}|
    \end{split}
\end{align}
where, in analogy to $P^{(1)}(u, w;v)$ above, $P^{(2)}(u-x, y-w;v)$ denotes the set of all shortest paths in the second-order network that connect
node $u-x$ to $y-w$ and that pass through a first-order node $v$.

With this, we have defined a second-order betweenness centrality which allows to calculate node centralities in a way that incorporates the causal topology as captured by the second-order aggregate network.
Let us again illustrate this approach using the simple examples shown in Fig.~\ref{fig:example}.
For the temporal network $G_1$ we can compute a second-order betweenness centrality based on the second-order network shown in Fig.~\ref{fig:example2nd:G1}.
Here we observe a total of four shortest paths between pairs of nodes in the second-order network, namely:
\begin{align*}
  (a-c, c-e) \\
  (a-c, c-d) \\
  (b-c, c-d) \\
  (b-c, c-e)
\end{align*}
For each node in the first-order network, we can now count the number of second-order shortest paths that they are on, obtaining $B^{(2)}(c) = 4$ while $B^{(2)}(x) = 0$ for all nodes $x \neq c$.
In this particular case, the second-order betweenness centrality values exactly correspond both to the temporal as well as the first-order betweenness centralities.
Again, this is different for the temporal network $G_2$ shown in Fig.~\ref{fig:example:G2}.
Considering the second-order aggregate network shown in Fig.~\ref{fig:example2nd:G2}, we only find the following two shortest paths in the second-order aggregate network
\begin{align*}
  (b-c, c-d) \\
  (a-c, c-e)
\end{align*}
thus obtaining $\text{BC}^{(2)}(c) = 2$.
Here, we find that while the second-order betweenness centralities in $G_2$ corresponds to the temporal betweenness centralities, they differ from those calculated from the first-order aggregate network.
The reason for this is that in the example $G_2$ shortest time-respecting paths of length two differ from what we would expect based on the first-order network.

We emphasize that the exact correspondence between the second-order and the temporal betweenness centralities in the examples discussed above is because we have no shortest time-respecting paths of length three or longer, whose presence could differ from what we expect based on the second-order network.
To what extent this affects the applicability of second-order aggregate networks in real-world scenarios is not clear and thus requires a further investigation.
In the following, we thus study to what extent second-order betweenness centrality can be used to approximate the temporal betweenness centralities of nodes in the six real-world data sets studied above.
For this, we first construct a second-order aggregate network as introduced in Section~\ref{sec:higherOrder}.
We then calculate the betweenness centrality values $\text{BC}^{(2)}(v)$ of all nodes $v$ as described above, comparing the resulting centralities with the (ground-truth) temporal betweenness centralities \\ $\text{BC}^{\text{temp}}(v)$.

The results of this analysis are shown in the right column of Table~\ref{tab:betweenness}.
Here we find that for most of the data sets, second-order betweenness centralities are correlated with the true, temporal betweenness centralities in a stronger way than the corresponding first-order approximation of betweenness centrality.
For the (EM) data sets capturing E-Mail exchanges between employees in a manufacturing company, we observe an increase of the Pearson correlation coefficient $\rho$ from $0.80$ to $0.97$, while the associated Kendall-Tau rank correlation coefficient $\tau$ increases rather mildly from $0.73$ to $0.74$.
We attribute this to the fact that the second-order aggregate network better captures the structures of time-respecting paths in the temporal network compared to the first-order network.
For the two data sets (HO) and (LT) we observe a similar increase both in the Pearson and the Kendall-Tau rank correlation coefficients, while the values remain largely unchanged for the (FL) data set.
In particular, for the latter data set the first-order betweenness centrality already exhibits a correlation coefficient of $0.99$ which indicates that in this particular case temporal characteristics do not significantly alter the structure of shortest time-respecting paths.
For the two data sets (AN) and (RM) we observe a small decrease in the Pearson correlation values for the second-order approximation.
Notably, for (RN) the decrease from $0.97$ to $0.95$ is accompanied by an increase of the Kendall-Tau coefficient from $0.62$ to $0.75$.
This indicates that, even though the actual values of second-order betweenness centralities may be less correlated with temporal betweenness centralities than the first-order betweenness centralities, the second-order betweenness centralities provides us with a significantly better perspective on the \emph{relative} importance of nodes.

Finally, for the (AN) data set we note that both the Pearson and the Kendall-Tau rank correlation coefficients are worse for the second-order betweenness centralities.
While the interesting question in what respect the temporal characteristics of (AN) differ from those of the other temporal networks remains to be investigated in more detail, we expect this result to be related to non-stationary properties.
We particularly observe that some of the nodes (i.e. ants) are only active during certain phases of the observation period.
This imposes a natural ordering of interactions which particularly prevents nodes which are only active during an early phase to be reachable from nodes which are only active at a later phase.

\subsection{Temporal Closeness Centrality}

Let us now turn our attention to \emph{closeness centrality}, which captures a node's average distance to all other nodes in a network.
For a directed, static (first-order aggregate) network the closeness centrality of a node $v$ is commonly defined as
\begin{align}
  \text{ClC}^{(1)}(v) = \sum_{u \neq v}{\frac{1}{\text{dist}^{(1)}(u,v)}}
  \label{eq:closeness:1st}
\end{align}
where the distance function $\text{dist}^{(1)}(u,v)$ denotes the distance, i.e. the length of a shortest path, from node $u$ to $v$ in the first-order aggregate network.

We can easily define a temporal version of closeness centrality based on the temporal distance function \\ $\text{dist}^{\text{temp}}(u,v)$ which we have defined in Eq.~\ref{eq:distance:temp} in the context of temporal betweenness centrality.
Here, we remind the reader that the function $\text{dist}^{\text{temp}}(u,v)$ captures the minimum length of a shortest time-respecting path across all possible start times $t_0$.
Using this temporal distance function, we can apply the standard definition in Eq.~\ref{eq:closeness:1st} and define the \emph{temporal closeness centrality} of a node $v$ in a temporal network as
\begin{align}
    \text{ClC}^{\text{temp}}(v) = \sum_{u \neq v} \frac{1}{\text{dist}^{\text{temp}}(u,v)}
\end{align}

Let us again illustrate this definition using the temporal networks shown in Fig.~\ref{fig:example}.
Node $e$ in the temporal network $G_1$ shown in Fig.~\ref{fig:example:G1} can be reached from nodes $a$ and $b$ via two shortest time-respecting paths of length two, as well as from node $c$ via a shortest time-respecting path of length one.
For the temporal closeness centrality, we thus find $\text{ClC}^{\text{temp}}(e) = 2$.
It is easy to confirm that this corresponds to the first-order closeness centrality of node $e$.
Again a mere reordering of links can change the closeness centralities of nodes, as can be seen in the temporal network $G_2$ shown in Fig.~\ref{fig:example:G2}.
Here, we see that node $e$ can only be reached from node $a$ via a shortest time-respecting path of length two, as well as from node $c$ via a shortest time-respecting path of length one.
For node $e$ in the temporal network $G_2$ we thus find a temporal closeness centrality $\text{ClC}^{\text{temp}}(e) = 1.5$, highlighting that it is, at least from the perspective of closeness centrality, less ``important'' than in the temporal network $G_1$.

Considering the example above we see that, due to the ordering and timing of links, first-order closeness centralities can be a misleading proxy for the temporal closeness centralities of nodes in temporal networks.
In the following we thus again empirically study this question using our six data sets on temporal networks.
We again use the temporal closeness centralities $\text{ClC}^{\text{temp}}(v)$ of nodes as the ground truth, then studying whether temporal closeness centralities can reasonably be approximated by first-order closeness centralities $\text{ClC}^{(1)}(v)$.
The results of this analysis are shown in the left column of Table~\ref{tab:closeness}, which reports the observed Pearson and Kendall-Tau rank correlation coefficients for each of the six data sets.
\begin{table}[t]
\centering
\begin{tabular}{|l|l|l|l|l|}
  \hline
                &  \multicolumn{2}{|c|}{$\text{ClC}^{\text{temp}} \thicksim \text{ClC}^{(1)}$}  &  \multicolumn{2}{|c|}{$\text{ClC}^{\text{temp}} \thicksim \text{ClC}^{(2)}$}    \\
                    & Pearson           & Kendall-Tau     & Pearson           & Kendall-Tau  \\
  \hline
  E-Mail (EM)       & 0.93 (4.74e-44)   & 0.79 (4.96e-30)  & 0.98 (2.52e-71)   & 0.92 (1.54e-40) \\
  Ants (AN)         & 0.91 (1.67e-24)   & 0.75 (1.54e-17)  & 0.96 (2.05e-35)   & 0.83 (4.80e-21) \\
  Hospital (HO)     & 0.96 (2.09e-29)   & 0.83 (1.88e-18)  & 0.99 (1.46e-40)   & 0.90 (1.76e-21) \\
  RealityMining (RM)& 0.96 (1.03e-33)   & 0.77 (1.99e-17)  & 0.99 (1.64e-51)   & 0.89 (5.30e-17) \\
  London Tube (LT)  & 0.98 (1.33e-91)   & 0.87 (2.57e-49)  & 0.98 (3.26e-92)   & 0.87 (1.07e-49) \\
  Flights (FL)      & 0.91 (3.35e-46)   & 0.81 (1.88e-18)  & 0.97 (4.57e-75)   & 0.93 (9.57e-50) \\
  \hline
\end{tabular}
\caption{Pearson and Kendall-Tau rank correlation coefficients between temporal closeness centrality (ground truth) and closeness centrality calculated based on the first-order aggregate network and the second order aggregate network. Values in parentheses indicate the p-value.}
\label{tab:closeness}
\end{table}

We observe again that the answer to the question of how well temporal closeness centralities can be approximated by first-order static closeness centralities depends on the actual data set.
The lowest Pearson correlation coefficient of $0.91$ is obtained for the (FL) and the (AN) data sets, while the highest Pearson correlation coefficient of $0.98$ is obtained for (LT).
The lowest Kendall-Tau rank correlation coefficient is $0.75$ for (AN), while the highest value of $0.87$ is achieved for (LT).
We further observe that, compared to betweenness centralities, we generally obtain conceivably larger correlation values between temporal and first-order closeness centralities.
This can intuitively be explained by the fact that, while temporal betweenness centralities are influenced by the actual \emph{structure} of shortest time-respecting paths, temporal closeness centralities are merely influenced by their lengths.
We thus expect temporal closeness centrality to be insensitive to characteristics of temporal networks that change the structure of paths but not their lengths, hence explaining the larger correlation coefficients.

Let us now study whether we can better approximate temporal closeness centralities using a generalization which is calculated based on the static, second-order aggregate representation of a temporal network.
For this we first introduce how closeness centralities of nodes can be calculated based on a second-order aggregate network.
We recall that in Eq.~\ref{eq:distance:2nd} we have defined a second-order distance function $\text{dist}^{(2)}(v,w)$ which provides us with the distance between (first-order) nodes based on shortest paths in a second-order aggregate network.
This distance function allows us to directly define a \emph{second-order closeness centrality} $\text{ClC}^{(2)}(v)$ as
\begin{align}
  \text{ClC}^{(2)}(v) = \sum_{u \neq v} \frac{1}{\text{dist}^{(2)}(u,v)}
  \label{eq:closeness:2nd}
\end{align}
i.e. for each node $v$ in a network, we simply sum the inverse of the distances to all nodes according to the topology of the second-order aggregate network.

Again, we illustrate the notion of second-order closeness centrality using the two illustrative examples of temporal networks shown in Fig.\ref{fig:example}.
Fig.~\ref{fig:example2nd:G1} shows the second-order aggregate network corresponding to the temporal network $G_1$ shown in Fig.~\ref{fig:example:G1}.
Here we find that the second-order node $c-e$ can be reached via two shortest paths
\begin{align*}
  (b-c),(c-e) \\
  (a-c),(c-e)
\end{align*}
of length one from the second-order nodes $b-c$ and $a-c$.
Furthermore, we have an additional second-order ``path'' of length zero from node $c-e$ to itself.
Using the second-order distance function as defined in Eq.~\ref{eq:distance:2nd}, we thus infer the following values:
\begin{align*}
  \text{dist}^{(2)}(b,e) = 2 \\
  \text{dist}^{(2)}(a,e) = 2 \\
  \text{dist}^{(2)}(c,e) = 1
\end{align*}
from which we calculate the second-order closeness centrality of node $e$ as $\text{ClC}^{(2)}(c) = 2$.

Again, in this particular example the second-order closeness centrality corresponds both to the temporal and the first-order closeness centrality.
This is different in the second-order network shown in Fig.~\ref{fig:example2nd:G2}, which corresponds to the temporal network $G_2$ shown in Fig.~\ref{fig:example:G2}.
Here, we find that the second-order node $c-e$ can only be reached via a single shortest path $(a-c),(c-e)$ as well as via an additional second-order ``path'' of length zero from $e-c$ to itself.
From this, we can calculate the following second-order distances
\begin{align*}
  \text{dist}^{(2)}(a,e) = 2 \\
  \text{dist}^{(2)}(c,e) = 1
\end{align*}
and for the second-order closeness centrality of node $e$ we thus obtain $\text{ClC}^{(2)}(c) = 1.5$, which coincides with the temporal closeness of node $e$ in the underlying temporal network $G_2$.

Using the the second-order closeness centrality introduced above, let us now study the correlations between the temporal and the second-order closeness centralities of nodes in our six data sets.
The results of this analysis are shown in the right column of Table~\ref{tab:closeness}.
For five of the six data sets we observe significantly larger correlation coefficients than those reported for the first-order closeness centrality in Table~\ref{tab:closeness}.
The largest increase of the Pearson correlation coefficient from $0.91$ to $0.97$ is achieved for the (FL) data set, while we observe no improvement of the (already large) Pearson correlation coefficient of $0.98$ for (LT).
We further observe significant increases in the Kendall-Tau rank correlation coefficients for all of the studied data sets, except for (LT) for which it remains the same.
For the ranking of nodes in (EM), we find that a ranking based on second-order closeness centralities increases the Kendall-Tau rank correlation with the ground truth temporal centralities from $0.79$ to $0.92$, thus better representing the relative importance of nodes in the temporal network.

\pagebreak

\subsection{Temporal Reach Centrality}

Concluding this section we finally study \emph{reach centrality}, another notion of path-based centrality that captures the number of nodes that can be reached from a node via paths up to given maximum length $s$~\cite{borgatti2013analyzing}.
For static networks, such as a first-order aggregate network, we define the \emph{first-order} reach centrality of a node $v$ as
\begin{align}
    \text{CoC}^{(1)}(v,s) := \sum_{w \in V} \Theta(\text{dist}^{(1)}(v,w)-s)\,
    \label{eq:reach:1st}
\end{align}
where $\Theta(\cdot)$ is the Heaviside function, $\text{dist}^{(1)}(v,u)$ is the length of a shortest path from node $v$ to $u$ in the static, first-order network, and $s$ is a parameter specifying up to which length paths should be considered.
Clearly, the reach centrality $\text{CoC}^{(1)}(v,s=1)$ of a node $v$ is equal to its out-degree while $\text{CoC}^{(1)}(v,s=\infty)$ is equal to the subset of nodes to which $v$ is connected via directed paths of any length.

A \emph{temporal reach centrality} can again easily be defined based on the notion of shortest time-respecting paths, as well as the temporal distance function $\text{dist}^{\text{temp}}(v,w)$ defined in Eq.~\ref{eq:distance:temp}.
Here, for a given maximum time difference $\delta$ 
and a given value $s$, we are interested in how many different nodes can be reached via shortest time-respecting paths which have at most length $s$.
In analogy to Eq.~\ref{eq:reach:1st}, we can thus define the \emph{temporal reach centrality} $\text{CoC}^{\text{temp}}(v)$ of a node $v$ as:
\begin{align}
    \text{CoC}^{\text{temp}}(v,s) := \sum_{w \in V} \Theta(\text{dist}^{\text{temp}}(v,w)-s).
    \label{eq:reach:temp}
\end{align}
We want to highlight that with this definition of reach centrality, we focus on the temporal-topological characteristics introduced by the ordering of links, which is why base our definition on the \emph{shortest} rather than the \emph{fastest} time-respecting paths.

It is finally easy to see that a \emph{second-order reach centrality} can be defined in analogy to second-order closeness centrality.
For this, all we have to do is to replace the distance function in Eq.~\ref{eq:reach:1st} by our previously defined second-order distance function, thus obtaining the following definition:
\begin{align}
    \text{CoC}^{(2)}(v,s) := \sum_{w \in V} \Theta(\text{dist}^{(2)}(v,w)-s)\, .
    \label{eq:reach:2nd}
\end{align}

Using a value of $s=2$, we again exemplify these definitions using our two illustrative examples.
Let us first calculate the first-order reach centrality of node $a$ based on the first-order aggregate network shown in Fig.~\ref{fig:example:agg}.
Here we find that there are paths of at most length $s=2$ from node $a$ to the three nodes $c,d$ and $e$, from which we conclude $\text{CoC}^{(1)}(a,s=2)=3$.
For the temporal reach centrality of node $a$ in the temporal network $G_1$ shown in Fig.~\ref{fig:example:G1}, we observe that there are time-respecting paths of at most length $s=2$ from node $a$ to the three nodes $c,e$ and $d$.
We hence conclude $\text{CoC}^{\text{temp}}(a,s=2)=3$, finding that for $G_1$ the temporal reach centrality again corresponds to the first-order reach centrality.
Again, this is not the case for the temporal network $G_2$ shown in Fig.~\ref{fig:example:G2}.
Here, node $a$ is only connected to the nodes $c$ and $e$ via time-respecting paths of up to length two, which means that we have $\text{CoC}^{\text{temp}}(a,s=2)=2$.

For the second-order reach centrality of node $a$ in the temporal network $G_1$ let us now consider the second-order aggregate network shown in Fig.~\ref{fig:example2nd:G1}.
Based on the shortest paths in the second-order network, we first find that the node $a-c$ is connected to two nodes $c-d$ and $c-e$ via shortest paths of length one.
Furthermore, we find an additional shortest path of length zero which connects the second-order node $a-c$ to itself.
Again, using our second-order distance function $\text{dist}^{(2)}$ here we find the distances
\begin{align*}
  \text{dist}^{(2)}(a,c) = 1 \\
  \text{dist}^{(2)}(a,e) = 2 \\
  \text{dist}^{(2)}(a,d) = 2
\end{align*}
from which we conclude that three nodes $c,e$ and $d$ can be reached via paths of length at most two.
From this we calculate the second-order reach centrality of node $a$ in $G_1$ as $\text{CoC}^{(2)}(a,s=2)=3$.
Applying the same arguments to the example network $G_2$ and the corresponding second-order aggregate network shown in Fig.~\ref{fig:example2nd:G2}, for the same three nodes we find the following second-order distances:
\begin{align*}
  \text{dist}^{(2)}(a,c) &= 1 \\
  \text{dist}^{(2)}(a,e) &= 2  \\
  \text{dist}^{(2)}(a,d) &= \infty
\end{align*}
We thus obtain a second-order reach centrality of \\ $\text{CoC}^{(2)}(a,s=2)=2$ which corresponds to the temporal reach centrality of node $a$ in $G_2$.

In the following, we use the temporal reach centrality defined above as ground truth, while studying how well it can be approximated by first-order and second-order reach centralities calculated from the first- and second-order time-aggregated networks respectively.
Different from the analyses for betweenness and closeness centralities, here we must additionally account for the fact that the reach centrality can be calculated for different values of the maximum path length $s$.
This implies that the Pearson correlation coefficient $\rho$ and the Kendall-Tau rank correlation coefficient $\tau$ must be calculated for each value of $s$ individually.
The results of this analysis are shown in Fig.~\ref{fig:reach}, which shows the obtained values for $\rho$ and $\tau$ for the correlations between i) the temporal and the first-order reach centralities (black lines), and ii) the temporal and the second-order reach centralities (orange lines) for each of the six data sets introduced above.
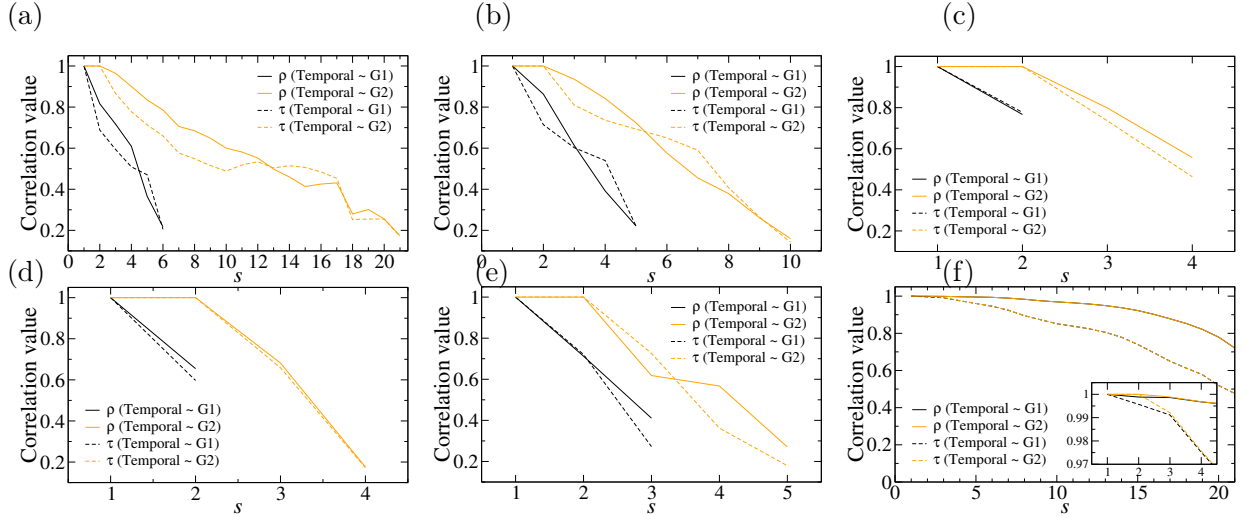
\begin{figure}[ht]
  \includegraphics[width=.32\textwidth]{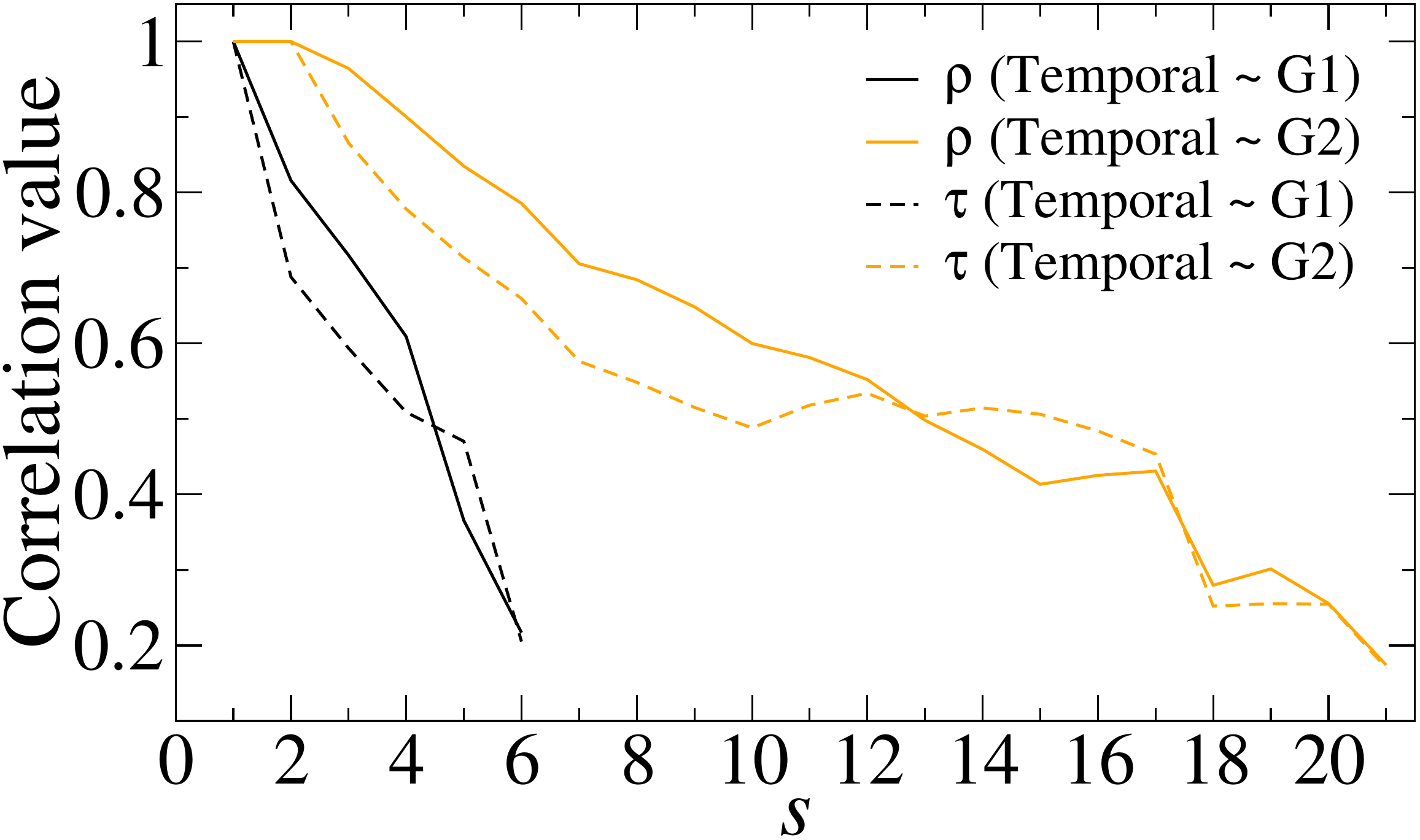}\hfill
  \includegraphics[width=.32\textwidth]{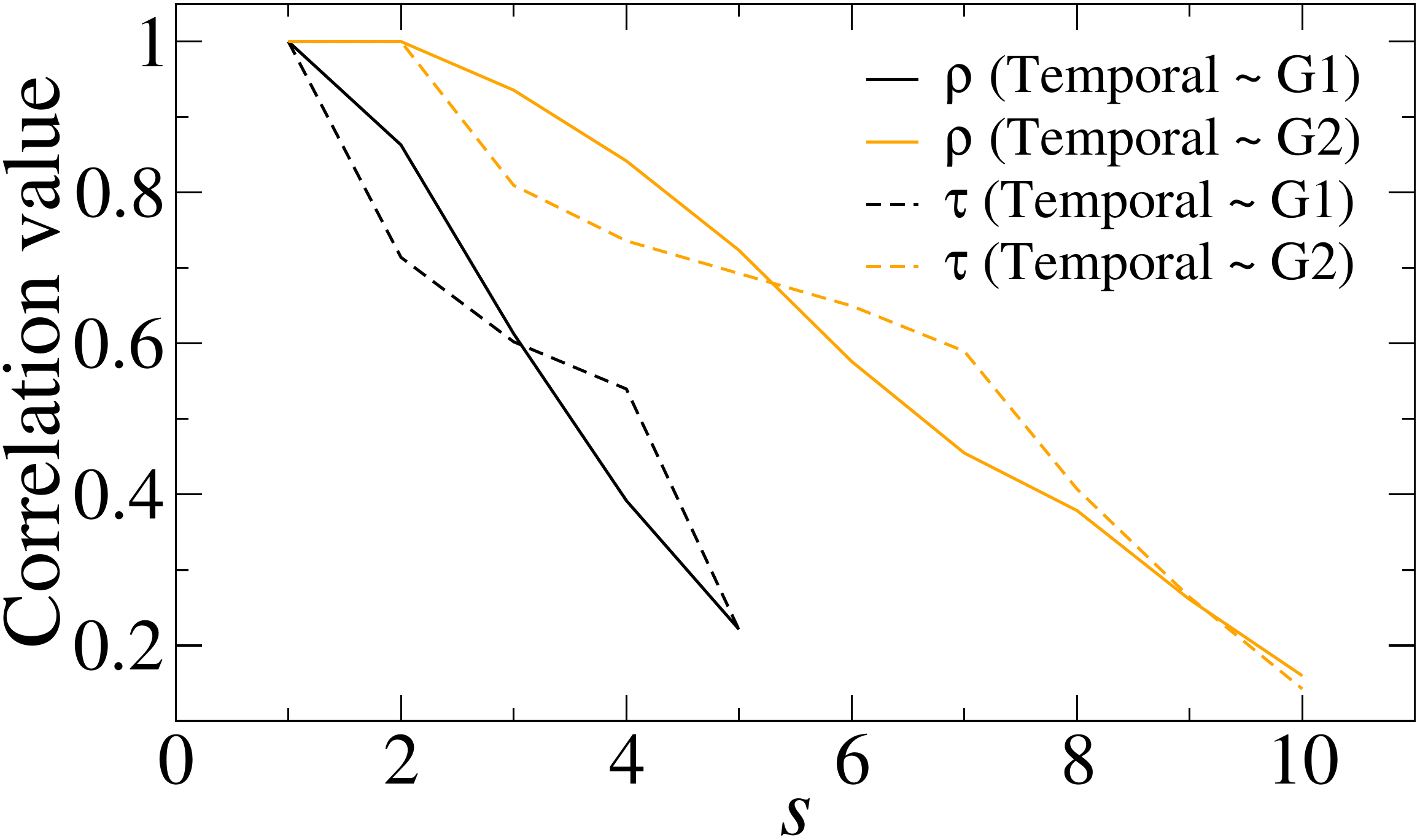}\hfill
  \includegraphics[width=.32\textwidth]{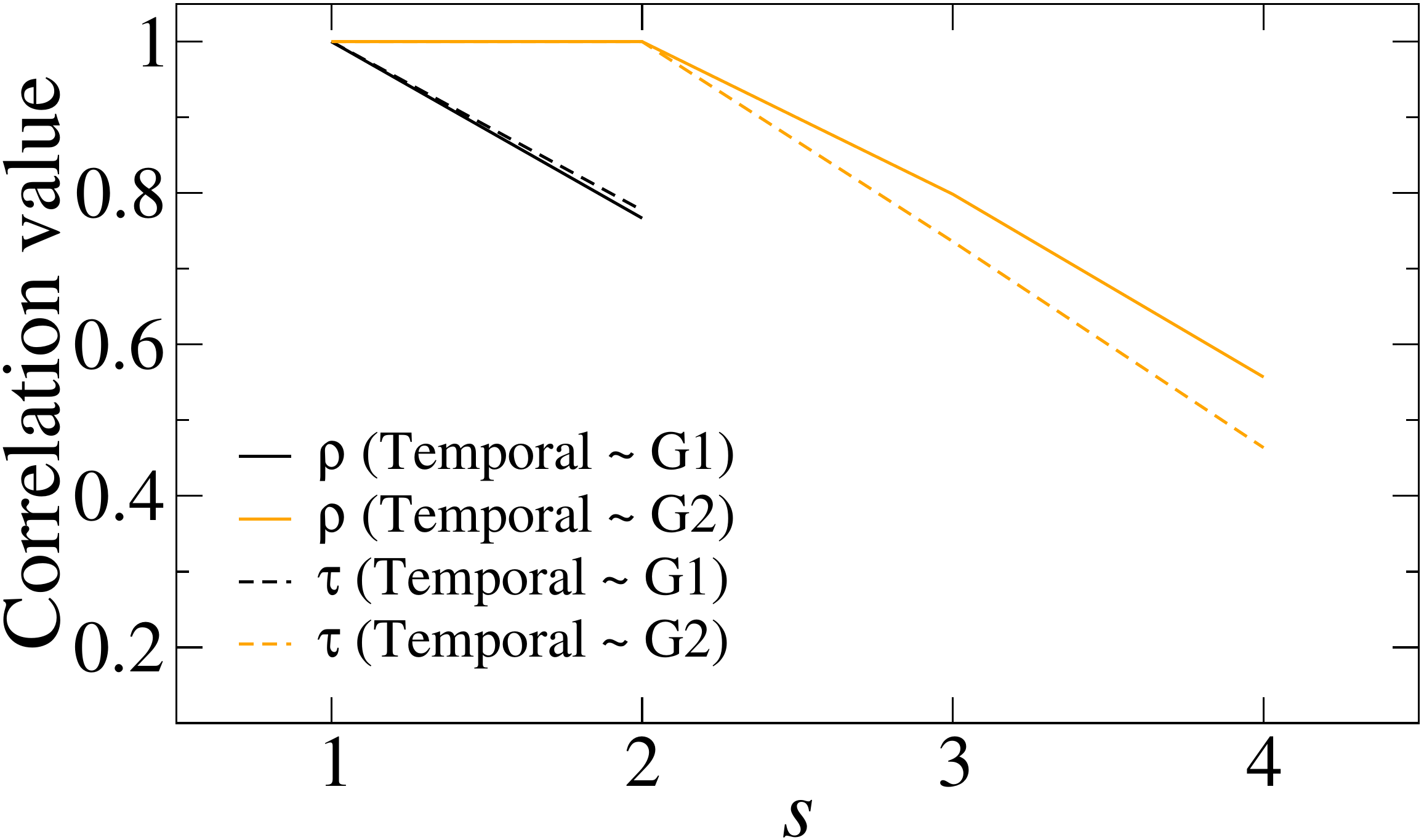}\\
  \includegraphics[width=.32\textwidth]{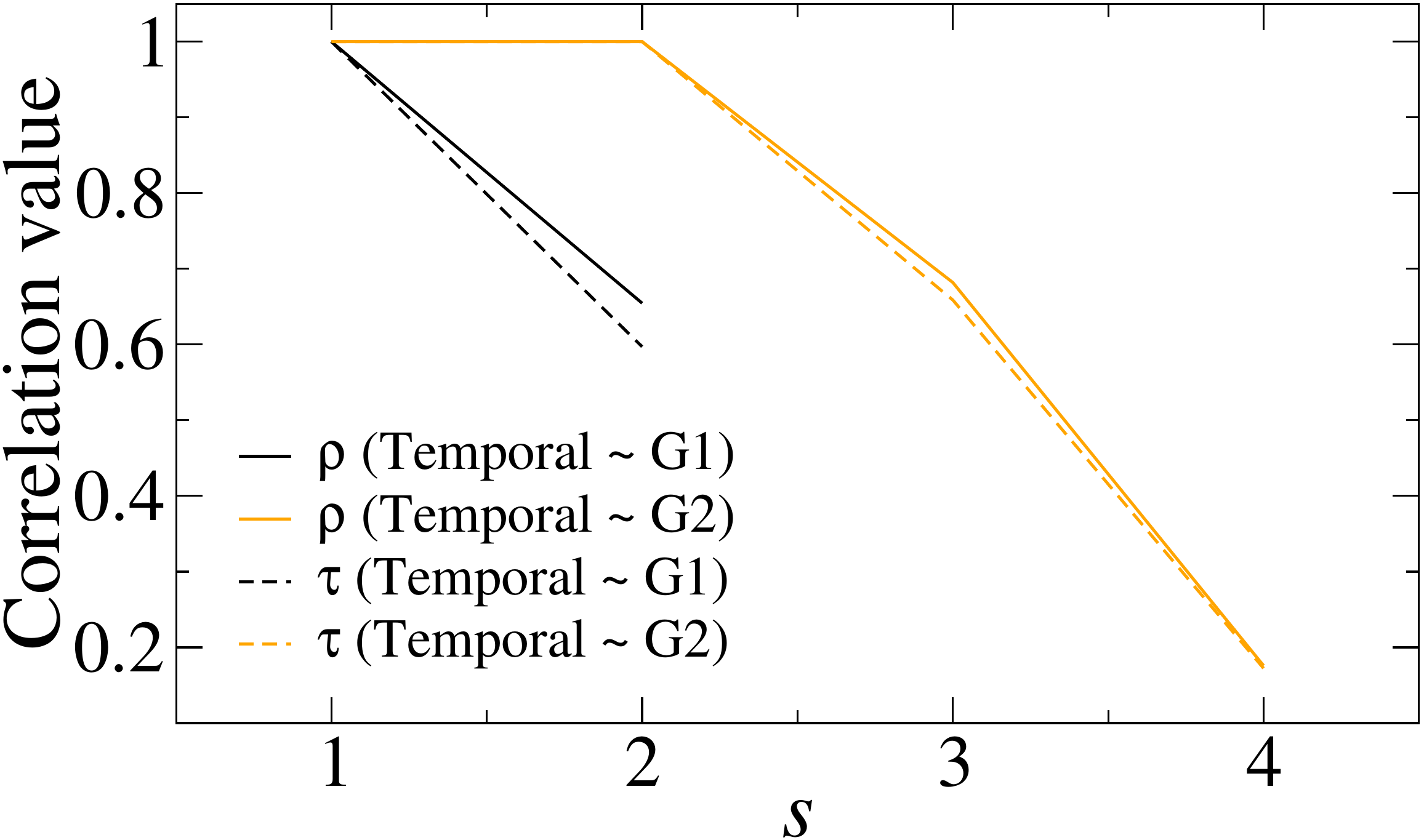}\hfill
  \includegraphics[width=.32\textwidth]{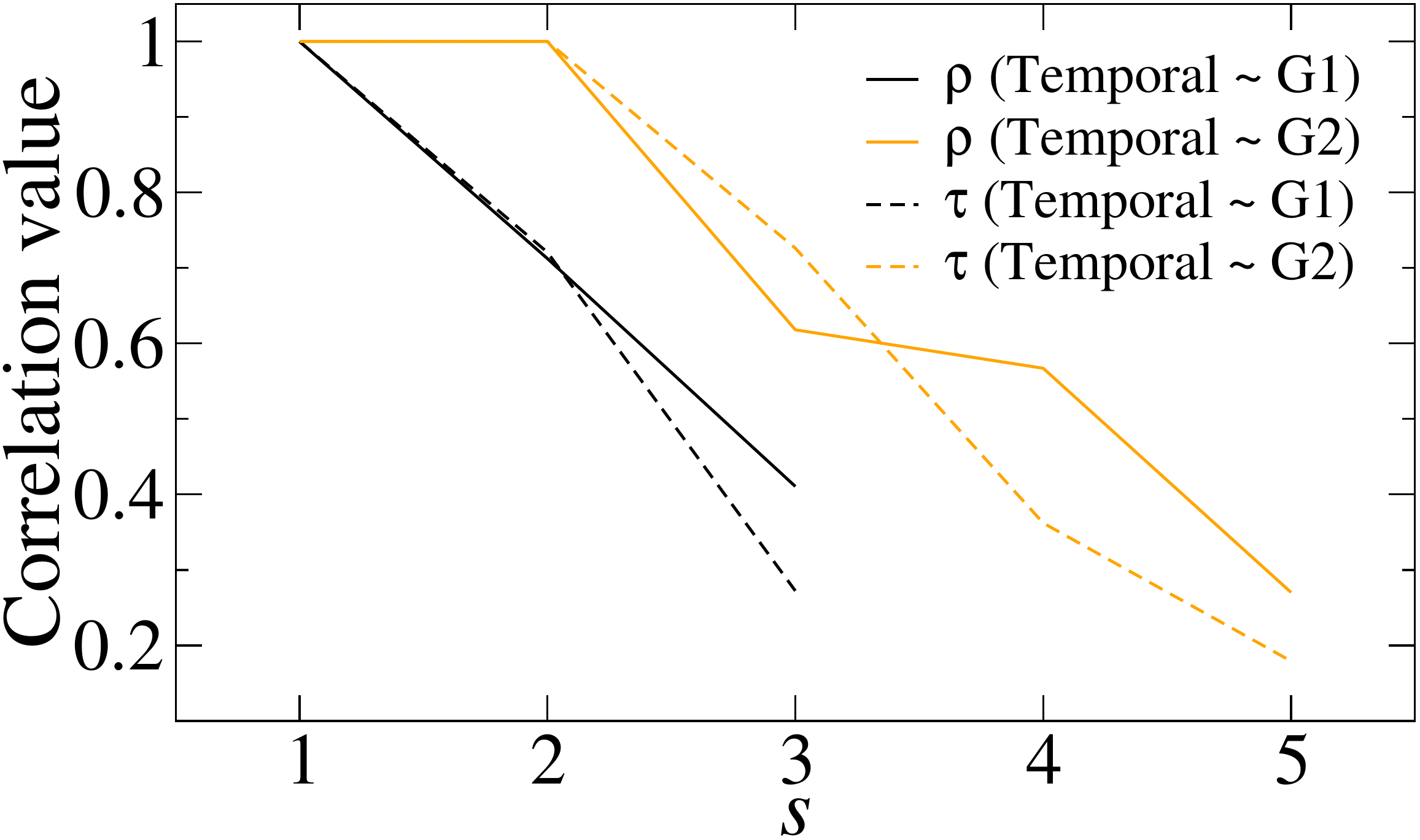}\hfill
  \includegraphics[width=.32\textwidth]{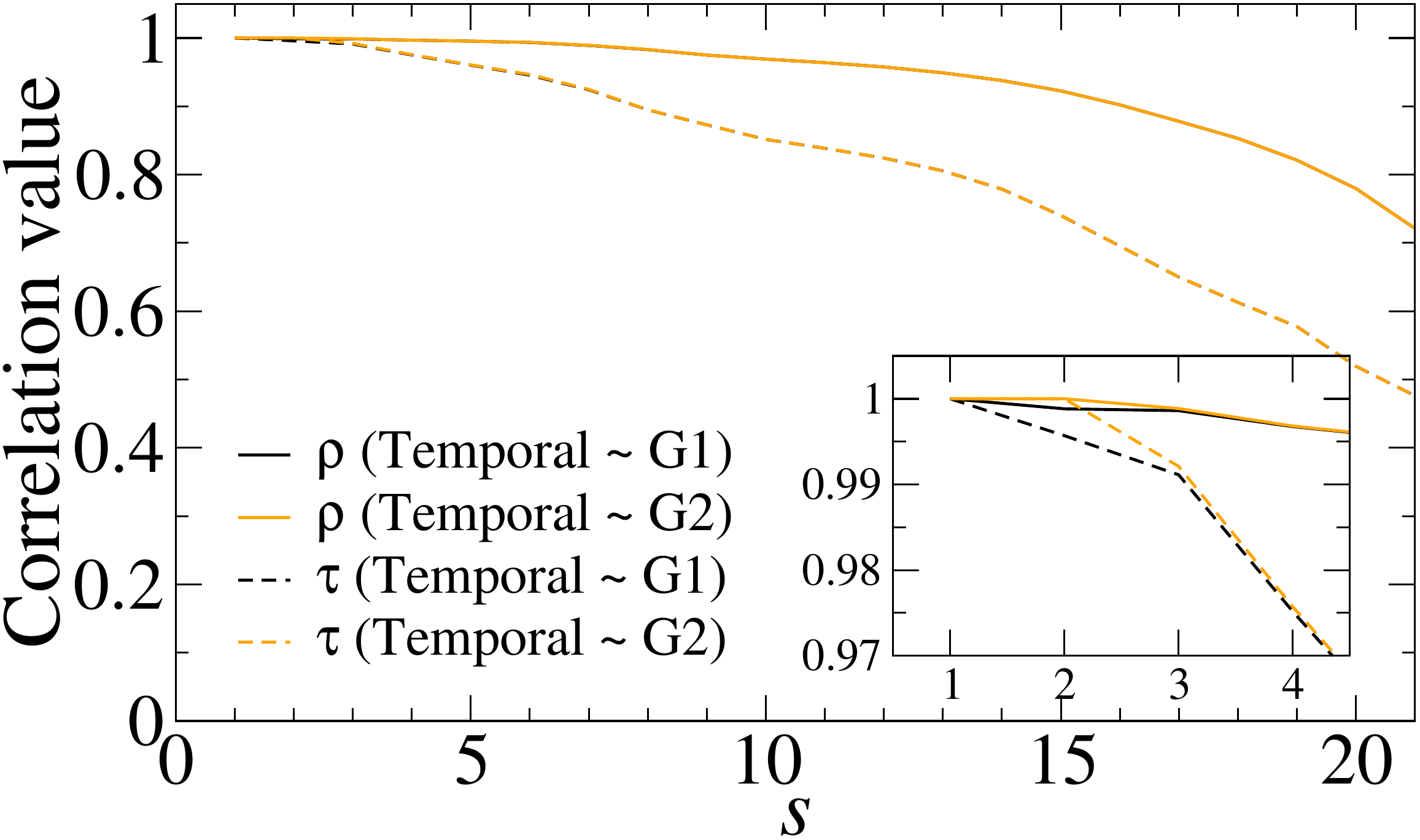}\\
  \begin{picture}(0,0)
    \put(-5,200){(a)}
    \put(170,200){(b)}
    \put(345,200){(c)}
    \put(-5,103){(d)}
    \put(170,103){(e)}
    \put(345,103){(f)}
  \end{picture}
  \caption{Pearson $\rho$ and Kendall $\tau$ correlation coefficients between the temporal and the first-order reach centralities (black lines) and the temporal and the second-order reach centralities (orange lines) for (a) the Ants data set, (b) the E-Mail data set, (c) the Hospital data set, (d) the Reality Mining dataset, (e) the Flights data set, and (f) the London Tube data set. Inset: zoom to the area where there is a small deviation between values for the case of the London Tube data set. }
  \label{fig:reach}
\end{figure}
Thanks to our choice of the maximum time difference $\delta$, for all of our data sets both the underlying first- and second-order networks are strongly connected.
Assuming that $D$ is the diameter of the corresponding aggregate network, for all $s \geq D$ we thus necessarily arrive at a situation where the reach centralities of all nodes are identical.
For the results in Fig.~\ref{fig:reach} this implies that for any $s>D$ the correlation values are undefined since the first- (or second-)order centralities of all nodes are the same.
We thus only plot the correlation coefficients $\tau$ and $\rho$ for $s<D$, in which case they are well-defined.

For $s=1$, the only time-respecting paths considered consist of single links, and thus the temporal reach centralities by definition exactly correspond to the reach centralities calculated from the first- and second-order topologies.
Consequently, for $s=1$ we have $\tau=1$ and $\rho=1$ both for the first- and the second-order reach centrality.
For $s=2$ there is, again by definition, no difference between the temporal and the second-order reach centralities however the correlation values for the first-order reach centrality decreases since the first-order aggregate network does not accurately represent the structure of time-respecting paths of length two.
For values $s>2$, $\rho$ and $\tau$ decrease both for the first and the second-order centralities since neither representation can accurately represent time-respecting paths with lengths $s>2$.
However the results also highlight the important fact that second-order reach centralities better approximate temporal reach centralities for all values of $s>2$.


We conclude this section by providing detailed results for the specific value of $s=3$.
The choice of a parameter $s>2$ means that for the second-order reach centrality we will not trivially obtain correlation values of $1$ because we would only consider time-respecting paths of length two which are captured in the second-order aggregate network.
However, since the diameter of the first-order aggregate network for two of our systems (RM and HO) is equal to three, we can only report results on the correlations between the temporal and the first-order reach centralities for four data sets.
The results for the first-order reach centrality with $s=3$ are shown in in Table~\ref{tab:reach}.
\begin{table}[ht]
\centering
\begin{tabular}{|l|l|l|l|l|}
  \hline
                &  \multicolumn{2}{|c|}{$\text{CoC}^{\text{temp}} \thicksim \text{CoC}^{(1)}$}  & \multicolumn{2}{|c|}{$\text{CoC}^{\text{temp}} \thicksim \text{CoC}^{(2)}$} \\
                & Pearson           & Kendall-Tau      & Pearson           & Kendall-Tau    \\
  \hline
  London Tube (LT)  & 1.00 (4.65e-168)   & 1.00 (9.32e-64)  & 1.00 (1.92e-173)  & 1.00 (7.00e-64)   \\
  Ants        (AN)  & 0.72 (8.23e-11)   & 0.59 (1.38e-11)   & 0.96 (9.50e-36)   & 0.86 (6.40e-23)  \\
  E-Mail      (EM)  & 0.61 (3.17e-11)   & 0.60 (3.55e-18)   & 0.94 (2.74e-44)   & 0.81 (1.52e-31) \\
  RealityMining (RM) & NA   & NA  & 0.68 (3.76e-09)   & 0.66 (2.78e-13) \\
  Hospital     (HO) & NA   & NA & 0.80 (7.95e-13)   & 0.74 (7.23e-15)  \\
  Flights      (FL) & 0.41 (4.68e-06)   & 0.27 (1.46e-05)   & 0.62 (1.44e-13)   & 0.73 (6.53e-31)  \\
  \hline
\end{tabular}
\caption{Pearson and Kendall-Tau rank correlation coefficients between temporal reach centrality (ground truth) and reach centrality for $s=3$ calculated based on the first--order aggregate network and the second-order aggregate network. Values in parentheses indicate the p-value.}
\label{tab:reach}
\end{table}

Remarkably, for the (LT) data sets we observe a perfect correlation with the temporal reach centrality, which means that for this data set reach centralities are seemingly not affected by the temporal characteristics of the system.
This is different for (FL), for which we observe a small Pearson correlation of $\rho=0.41$, with an associated $\tau=0.27$.
These results show that, for the (FL) data set, temporal characteristics of the data do not allow temporal reach centralities to be approximated based on the first-order aggregate network.
For the second-order reach centralities shown in the right columns of Table~\ref{tab:reach}, we observe a significant increase in both the Pearson and the Kendall-Tau correlation coefficients for all of the data sets, except for (LT).
The largest increase of the Pearson correlation coefficient is again obtained for (EM), increasing from $0.61$ to $0.94$ with an associated increase of the Kendall-Tau correlation coefficient from $0.60$ to $0.81$.
We thus conclude that again, second-order reach centralities better capture the true (temporal) importance of nodes than a simple first-order approximation.

\section{Conclusion}
\label{sec:conclusion}

In summary, we have introduced a framework for the analysis of path-based notions of node centralities in temporal networks.
In particular, we defined temporal versions of three path-based centrality measures which highlight the influence of the temporal-topological dimension introduced by the specific timing and ordering of time-stamped links in temporal networks.
Using six data sets on real-world temporal networks, we have studied to what extent static notions of betweenness, closeness and reach centrality differ from their temporal counterparts.
While for some data sets node centralities in the (first-order) time-aggregated, static network can be used as reasonable proxies for temporal centralities, our results show that for other data sets this is not the case.
Here we found that an analysis of time-aggregated static networks that neglect the time dimensions can yield misleading results about the importance of nodes.

In order to overcome these limitations, we have further introduced higher-order aggregate networks, a simple yet powerful generalization of the commonly used time-aggregated static perspective on time-stamped network data.
The basic idea of this construction is that a $k$-th order aggregate network captures the statistics of time-respecting paths of length $k$, thus facilitating a higher-order analysis that incorporates both the topology and the ordering of links in temporal networks.
We demonstrate the power of this framework through the definition of \emph{second-order} centralities which can easily be calculated based on shortest paths in a second-order aggregate network.
Despite the fact that these centralities can easily be calculated based on a simple static network structure, we find that the resulting second-order centrality measures capture better the true temporal centralities of nodes in the underlying temporal networks.

Closing, we would like to highlight a number of open issues which we plan to consider in future works.
First and foremost, all of our results have been obtained based on simple \emph{unweighted} notions of centralities, even though in principle both the first-and second-order aggregate networks allow for the definition of link weights.
Hence, our results have been obtained based on a rather simple perspective which does not incorporate the full information about path statistics preserved by our higher-order aggregate network abstraction.
We thus expect a future extension to weighted higher-order aggregate networks to capture the true temporal centralities of nodes even more closely.
Furthermore, while we can in principle define higher-order networks of any order $k$, in our work we have merely studied second-order representations and the corresponding generalizations of path-based centralities.
Our choice to limit our study to an order of $k=2$ is mainly due to the amount of available data, which for the six temporal networks studied in this work does not allow to obtain meaningful statistics for time-respecting paths with larger lengths of size $k$ that are the basis for a $k$-th order aggregate network.
Under what conditions higher-order aggregate networks with orders of $k>2$ can help us to obtain even better approximations for temporal centralities is thus an open question that should be studied in the future.

Despite these open issues, we consider the fact that the simple second-order centrality measures introduced in our work already yield good approximations of the underlying temporal centralities a promising aspect of our framework.
In this respect, second-order time-aggregated representations of temporal networks can be considered a simple, yet powerful abstraction for the higher-order analysis of time-stamped network data.

\bibliographystyle{abbrv}
\bibliography{article}

\end{document}